\documentclass[conference]{IEEEtran}
\IEEEoverridecommandlockouts

\usepackage{amsmath,amsfonts}
\usepackage{algorithmic}
\usepackage{graphicx}
\usepackage{textcomp}
\usepackage{xspace}
\usepackage{xcolor}
\usepackage{cancel}
\usepackage{soul}
\usepackage{enumitem}
\setlist[itemize]{leftmargin=8mm}
\usepackage{url}
\usepackage[hidelinks]{hyperref}
\usepackage{threeparttable}
\usepackage{booktabs}
\usepackage{listings}
\usepackage{flushend}
\usepackage{svg}
\usepackage[normalem]{ulem}

\usepackage{tcolorbox}%
\definecolor{mycolor}{rgb}{0.122, 0.435, 0.698}
\definecolor{gray1}{gray}{0.3}

\definecolor{darkgreen}{rgb}{0.0, 0.5, 0.0}
\definecolor{darkred}{rgb}{0.82, 0.1, 0.26}

\newcommand{\result}[1]{%
\begin{tcolorbox}[colframe=mycolor,boxrule=0.5pt,arc=4pt,
      left=6pt,right=6pt,top=6pt,bottom=6pt,boxsep=0pt,width=\columnwidth]%
      {#1}
\end{tcolorbox}%
}

\newcommand{\sqlancer}{\textsl{SQLancer}\xspace}
\newcommand{\method}{\textsl{UPlan}\xspace}

\newcommand{\qpg}{\textsl{QPG}\xspace}
\newcommand{\cert}{\textsl{CERT}\xspace}
\newcommand{\mozi}{\textsl{Mozi}\xspace}

\newcommand{\numbugs}{17\xspace}

\newcommand{\bugsymbol}[0]{\includegraphics[width=0.3cm]{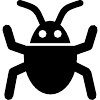}}
\newcommand{\oksymbol}[0]{\includegraphics[width=0.3cm]{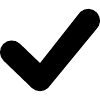}}

\renewcommand{\sectionautorefname}{Section}
\renewcommand{\subsectionautorefname}{Subsection}

\newcommand{\Autoref}[1]{%
  \begingroup%
  \def\chapterautorefname{Chapter}%
  \def\sectionautorefname{Section}%
  \def\subsectionautorefname{Subsection}%
  \autoref{#1}%
  \endgroup%
}

\def\ojoin{\setbox0=\hbox{$\bowtie$}%
  \rule[-.02ex]{.25em}{.4pt}\llap{\rule[\ht0]{.25em}{.4pt}}}
\def\leftouterjoin{\mathbin{\ojoin\mkern-5.8mu\bowtie}}

\newcommand{\etal}{\textit{et al}.}
\newcommand{\ie}{\textit{i}.\textit{e}.}
\newcommand{\eg}{\textit{e}.\textit{g}.}

\begin{document}

\title{Towards a Unified Query Plan Representation}

\author{
\IEEEauthorblockN{Jinsheng Ba\thanks{The first author started the project at the National University of Singapore.}}
\IEEEauthorblockA{ETH Zurich}

\and
\IEEEauthorblockN{Manuel Rigger}
\IEEEauthorblockA{National University of Singapore}
}

\maketitle

\begin{abstract}
In database systems, a query plan is a series of concrete internal steps to execute a query. Multiple testing approaches utilize query plans for finding bugs. However, query plans are represented in a database-specific manner, so implementing these testing approaches requires a non-trivial effort, hindering their adoption. We envision that a unified query plan representation can facilitate the implementation of these approaches. In this paper, we present an exploratory case study to investigate query plan representations in nine widely-used database systems. Our study shows that query plan representations consist of three conceptual components: operations, properties, and formats, which enable us to design a unified query plan representation. Based on it, existing testing methods can be efficiently adopted, finding 17 previously unknown and unique bugs. Additionally, the unified query plan representation can facilitate other applications. Existing visualization tools can support multiple database systems based on the unified query plan representation with moderate implementation effort, and comparing unified query plans across database systems provides actionable insights to improve their performance. We expect that the unified query plan representation will enable the exploration of additional application scenarios.
\end{abstract}


\section{Introduction}

Database Management Systems (DBMSs) are fundamental software systems used to store, retrieve, and run queries on data. They are used in almost every computing device~\cite{sqlitedeployed, tidbdeployed, cockdorachdbdeployed}; thus, any bug has potentially severe consequences. Recently, automated testing approaches for DBMSs have gained broad adoption. \emph{Query Plan Guidance} (\qpg)~\cite{qpg} guides test case generation towards diverse query plans aiming to expose more bugs, \emph{Cardinality Estimation Restriction Testing} (\cert)~\cite{cert} identifies performance bugs by examining query plans, and \mozi~\cite{liang2024mozi} checks the consistency across query plans of the same query to find incorrect query optimizations.

While these approaches have been effective and have found more than 200 previously unknown and unique bugs, implementing them requires a non-trivial effort, as they require DBMS-specific logic to parse and process query plans.
A query plan refers to a series of concrete steps to execute a query written in a declarative language, such as \emph{Structured Query Language} (SQL).
DBMSs expose query plans in various formats (\eg, in textual format), in which case we refer to them as \emph{serialized query plans}.
Unlike query languages, for which widely-used, standardized~\cite{paolo2018how}, and formalized~\cite{paolo2017a} languages exist, such as SQL~\cite{sqliso}, the formats in which serialized query plans are exposed are non-standardized and DBMS-specific. For example, a predicate in the \lstinline{WHERE} clause of SQL corresponds to a concrete step to filter data in the query plans of TiDB~\cite{tidb}, but corresponds to a property of another step to scan tables in the query plans of PostgreSQL~\cite{postgresql}. We refer to the different ways in which serialized query plans are represented as \emph{query plan representations}. 
Considering that hundreds of DBMSs exist,\footnote{\url{https://dbdb.io/} lists 1001 DBMSs as of August 2024.} implementing the above testing methods requires significant effort as they need to account for differences in query plan representations, thus significantly hindering the effectiveness of the above approaches.

We envision that a unified query plan representation would remove the roadblock to implementing the above testing approaches.
In this work, we systematically study query plan representations. We present an exploratory case study~\cite{host2012case}, which is a method to investigate a phenomenon in depth, including both qualitative and quantitative research methods. We collected documents, source code, and third-party applications of the query plans in nine popular DBMSs across five different data models, and summarized the commonalities and differences of query plan representations. Our study shows that query plan representations are based on three conceptual components: operations, properties, and formats. Based on the study, we designed a unified query plan representation, which allows the representation of all conceptual components we studied from DBMS-specific query plans, so that we can easily apply automated testing methods to multiple targets.

To demonstrate the utility of the unified query plan representation, we implemented a prototype called \method to maintain the representation. Currently, we implemented converters that convert a DBMS-specific serialized query plan to a unified representation; in the long term, we hope that DBMSs will directly expose such a unified representation, rendering such converters unnecessary. We re-implemented \qpg and \cert in a DBMS-agnostic way based on \method, enabling the large-scale adoption of both testing methods. In our evaluation, we considered three DBMSs: MySQL, PostgreSQL, and TiDB. We found \numbugs previously unknown and unique bugs, and one bug in the original \qpg implementation for parsing TiDB's query plans. Additionally, we found that \method can facilitate other applications based on query plans. First, we modified a visualization tool to support our unified query plan representation. We show that an existing DBMS-specific visualization tool can support five DBMSs through \method with only moderate implementation effort. Second, the unified query plan representation enables the comparison of query plans in different DBMSs, which provides actionable insights for query optimization. 

We believe that our unified representation reduces the effort to build applications, including testing, on serialized query plans. We include all study results and the prototype in our supplementary materials.\footnote{\url{https://zenodo.org/records/15097905}} For ease of access, we also provide a comprehensive website,\footnote{\url{https://nus-test.github.io/uplan/}} which provides supportive additional information including illustrative examples, explanations of studied query plans, and example applications. We make the following contributions:
\begin{itemize}
    \item A study of query plan representations and results that allow both practitioners and researchers to study the results.
    \item A proposal and a reusable library \method for a unified query plan representation.
    \item Three applications, including testing, of the unified representations on serialized query plans.
\end{itemize}
\section{Background}\label{sec:background}

\emph{Query optimization.} DBMSs consider a variety of potential execution strategies for a query written in a declarative language~\cite{ioannidis1996query}, and each execution strategy includes specific steps to execute the query. Query optimization is the process of determining an execution strategy, and typically includes these steps: parsing queries into logical query plans that represent the logical steps to execute the query, including operations such as join, converting them into physical query plans, which represent the executable operations to execute the query, such as hash join, and determining a physical query plan to execute. 

\emph{Query plans.} Query plans, typically short for physical query plans, are organized as a Directed Acyclic Graph (DAG)~\cite{neumann2005efficient}, in which a step outputs data to another step as input followed by directed edges. As query optimizations are specific to the storage engines and query execution engines, query plans are non-standardized. By executing a query with a specific prefix, such as \lstinline{EXPLAIN} for most relational DBMSs that support SQL, we can obtain a \emph{serialized query plan} in a DBMS-specific \emph{query plan representation}. In this paper, we study various query plan representations. We did not study logical query plans as they usually are not exposed by DBMSs.
\section{Query Plan Case Study}\label{sec:study}

We adopted an exploratory case study as the method to investigate query plan representations. The case study, as an empirical method, is used for investigating a contemporary phenomenon in depth and within its real-world context~\cite{yin2009case} (the ``case"). An exploratory case study is a specific case study that generates new questions, propositions, or hypotheses during the study. In this paper, we chose this method, because we wanted to gain an in-depth understanding of how query plans are represented within mature DBMSs. We followed common guidelines of case study research~\cite{host2012case} to design and conduct this study.

\subsection{Case Study Design}

\begin{table}[]
    \centering
    \caption{The studied nine DBMSs ranging from various data models, development modes, and release dates.}
    \begin{threeparttable}
        \begin{tabular}{@{}lllrr@{}}
            \toprule
            \textbf{DBMS} & \textbf{Version} & \textbf{Data Model} & \textbf{Release}  & \textbf{Rank}  \\
            \midrule
            InfluxDB~\cite{influxdb} & 2.7.0 & Time-series & 2013 & 28 \\
            MongoDB~\cite{mongodb} & 6.0.5 & Document & 2009 & 5 \\
            MySQL~\cite{mysql} & 8.0.32 & Relational & 1995 & 2 \\
            Neo4j~\cite{neo4j} & 5.6.0 & Graph & 2007 & 21 \\
            PostgreSQL~\cite{postgresql} & 14.7 & Relational & 1989 & 4 \\
            SQL Server~\cite{sqlserver} & 16.0.4015.1 & Relational & 1989 & 3 \\
            SQLite~\cite{sqlite} & 3.41.2 & Relational & 1990 & 10 \\
            SparkSQL~\cite{sparksql} & 3.3.2 & Relational & 2014 & 33 \\
            TiDB~\cite{tidb} & 6.5.1 & Relational & 2016 & 79 \\
            \bottomrule
        \end{tabular}
    \end{threeparttable}
    \label{tab:benchmark}
\vspace{-5mm}
\end{table}

\emph{Objectives and research questions.} The goal of this study was to investigate the phenomenon of non-standardized query plan representations within real-world DBMSs.
We aim to achieve this goal by answering the following research questions (RQs):

\begin{itemize}
    \item[RQ1] How are serialized query plans represented?
    \item[RQ2] Do query plan representations share a common conceptual basis?
\end{itemize}

\emph{Case selection.} The case study of this paper is characterized as single-case~\cite{host2012case}: the query plan representation is the case, while different DBMSs are the units of the analysis. \autoref{tab:benchmark} shows the DBMSs that we selected for the study. To conduct a representative and comprehensive study, we made a diverse selection of DBMSs. First, we chose DBMSs of various data models: relational, document, graph, and time-series data models, based on the assumption that the DBMSs of different models might have different query plan representations. The relational data model is the most widely-used one~\cite{e.1970a}, and other non-relational models, which are also called NoSQL models, are widely used for maintaining unstructured or semi-structured data~\cite{strauch2011nosql}. Then, we chose both classic and new-generation DBMSs~\cite{pavlo2016s} ranging from release years from the 1980s to the 2010s. The architectures of the chosen DBMSs include standalone DBMSs and an embedded DBMS, SQLite, which runs in the same process as the application. We included both a commercial DBMS, SQL Server, and open-source DBMSs. Apart from conventional DBMSs, we included an analytics engine for large-scale data processing, SparkSQL, which also optimizes queries. To choose widely-used DBMSs for study, we chose them referring to the DBMS ranking website\footnote{\url{https://db-engines.com/en/ranking} as of August 2024.} as shown in the column \emph{Rank}. We chose the latest release version of each DBMS.

\emph{Data collection.} We collected data from multiple sources: documents, source code, officially integrated development environments (IDEs), and third-party applications based on query plans. The use of multiple data sources allowed us to perform data source triangulation~\cite{host2012case}, that is, we could confirm the study results from the above four different types of data sources. We included the detailed lists of the data sources in the supplementary materials for reference.

\emph{Data analysis.} For each DBMS, we first examined its official documents describing its query plan representation. Where the source code was available, we inspected it to gain a better understanding of query plan representations. We also ran official test cases in the official IDEs to observe query plan representations in real-world test cases. For the DBMSs that had open-source query plan applications, we further examined how query plan representations are explained and utilized from a third-party perspective. To answer RQ1, we analyzed the query plan representations from the above data sources qualitatively. To answer RQ2, we qualitatively identified conceptual components in the query plan representations, and quantitatively compared them. To satisfy observation triangulation~\cite{host2012case}, one author conducted the study, and another author validated the finding against the raw data.

\begin{figure}[]
\begin{lstlisting}[caption={Query plan examples of PostgreSQL and SQLite in text format. Some properties are truncated by ... due to space constraints.},captionpos=t, label=lst:example, escapeinside=@@]
CREATE TABLE t0 (c0 INT);
CREATE TABLE t1 (c0 INT);
CREATE TABLE t2 (c0 INT PRIMARY KEY);
INSERT INTO t0 SELECT * FROM generate_series(1,1000000);
INSERT INTO t2 SELECT * FROM generate_series(1,100);

----------------------PostgreSQL-----------------------
EXPLAIN (SUMMARY TRUE) SELECT t1.c0 FROM t0 INNER JOIN t1 ON t0.c0 = t1.c0 WHERE t0.c0 < 100 GROUP BY t1.c0 UNION SELECT c0 FROM t2 WHERE c0 < 10;

@\textbf{HashAggregate}@ (cost=62998.82..63009.32 rows=1050...)
 Group Key: t1.c0
 ->@\textbf{Append}@ (cost=27150.40..62996.20 rows=1050 width=4)
   ->@\textbf{Group}@ (cost=27150.40..62949.08 rows=200 width=4)
     Group Key: t1.c0
     ->@\textbf{Gather Set}@ (cost=27150.40..62948.08 rows=400...)
       Workers Planned: 2
       ->@\textbf{Group}@ (cost=26150.38..61901.89 rows=200...)
         Group Key: t1.c0
         ->@\textbf{Set Join}@ (cost=26150.38..56906.48...)
           Set Cond: (t0.c0 = t1.c0)
           ->@\textbf{Sort}@ (cost=25970.60..26362.39...)
             Sort Key: t0.c0
             ->Parallel @\textbf{Seq Scan}@ on t0 (cost=0.00...)
               Filter: (c0 < 100)
           ->@\textbf{Sort}@ (cost=179.78..186.16 rows=2550...)
             Sort Key: t1.c0
             ->@\textbf{Seq Scan}@ on t1 (cost=0.00..35.50...)
   ->@\textbf{Bitmap Heap Scan}@ on t2 (cost=10.74..31.37...)
     Recheck Cond: (c0 < 10)
     ->@\textbf{Bitmap Index Scan}@ on t2_pkey (cost=0.00...)
       Index Cond: (c0 < 10)
Planning Time: 0.124 ms

-------------------------SQLite-------------------------
EXPLAIN QUERY PLAN SELECT t1.c0 FROM t0 INNER JOIN ...;

`--@\textbf{COMPOUND QUERY}@
   |--@\textbf{LEFT-MOST SUBQUERY}@
   |  |--@\textbf{SCAN}@ t0
   |  |--@\textbf{SEARCH}@ t1 USING AUTOMATIC COVERING INDEX (c0=?)
   |  `--@\textbf{USE TEMP B-TREE}@ FOR GROUP BY
   `--@\textbf{UNION}@ USING TEMP B-TREE
      `--@\textbf{SEARCH}@ t2 USING COVERING INDEX ...

\end{lstlisting}
\vspace{-5mm}
\end{figure}

\subsection{Findings Overview}
We found that the studied DBMSs share three conceptual components: \emph{operations}, \emph{properties}, and \emph{formats}. 
As described in \Autoref{sec:background}, query plans are considered DAGs. In practice, serialized query plans are usually organized in a tree structure, and each node has a key. The keys typically refer to \emph{operations}, which are concrete steps executed by DBMSs to retrieve, process, or output data in response to a query, such as \lstinline{Full Table Scan}, which refers to the step to scan an entire table. Each \emph{operation} is associated with zero or multiple \emph{properties}, which involve \emph{operation}-related information, such as \lstinline{row}, which refers to the estimated number of rows returned. Apart from \emph{operation}-associated \emph{properties}, plans also have \emph{properties} associated with them, such as \lstinline{planning time}, which refers to the time to generate the query plan. A special case is the key \emph{Filter} in the TiDB query plans. It represents the condition that the output data from its child node satisfies, so we deem it as a \emph{property} instead of an \emph{operation}. DBMSs typically allow serializing query plans in various \emph{formats}, such as text, table, \textsc{JSON}, and \textsc{XML}. 
 
\autoref{lst:example} shows two examples of query plan representations of PostgreSQL and SQLite in a textual format. Lines 1--5 show the SQL statements that create and populate the tables. The function \lstinline{generate_series} generates data to populate tables \lstinline{t0} and \lstinline{t2}. For PostgreSQL, executing the statement in line 8 outputs the serialized query plan as shown in lines 10--32. 
The bold texts denote operations, and the non-bold texts denote properties. For example, the operation \textbf{\footnotesize HashAggregate} in line 10 has the properties \lstinline{cost, rows, width, and Group Key}. The tree structure is denoted through hierarchical indents, in which the operation with longer indents is a child of the operation with short indents. Although both DBMSs are based on the relational data model, and support a textual format, their query plan representations are significantly different.

\begin{table*}
    \caption{The number of operations and properties in query plan representations. RA: Relational Algebra.}
    \label{tab:number}
    \centering\scriptsize
    \begin{tabular}{@{}l@{}rrrrrrrrrrrrr@{}}
        \toprule
         & \multicolumn{8}{c}{\textbf{Operations}} & \multicolumn{5}{c}{\textbf{Properties}} \\
         \cmidrule(lr){2-9} \cmidrule(lr){10-14}
        \textbf{DBMS}   & \textbf{Producer} & \textbf{Combinator} & \textbf{Join} & \textbf{Folder} & \textbf{Projector} & \textbf{Executor} & \textbf{Consumer} & \textbf{Sum} & \textbf{Cardinality} & \textbf{Cost} & \textbf{Configuration} & \textbf{Status} & \textbf{Sum} \\
        \textbf{RA Operators} & $\sigma$ & $\cup, \cap, -$ & $\bowtie, \times$ & $\gamma$ & $\Pi$ \\
        \midrule               
        InfluxDB	&0	&0	&0	&0	&0	&0	&0	& 0     &5	&0	&0	&1	&6 \\
        MongoDB	    &14	&9	&0	&5	&3	&10	&3	& 44    &16	&5	&18	&12	&51 \\
        MySQL	    &15	&3	&2	&1	&0	&2	&0	& 23    &3	&6	&3	&10	&22 \\
        Neo4j	    &18	&11	&43	&6	&3	&17	&13	& 111   &3	&3	&12	&7	&25 \\
        PostgreSQL	&18	&8	&3	&3	&0	&9	&1	& 42    &8	&17	&42	&40	&107 \\
        SQL Server	&15	&3	&3	&3	&0	&16	&19	& 59    &4	&4	&7	&3	&18 \\
        SQLite	    &3	&6	&3	&0	&0	&5	&0	& 17    &0	&0	&3	&0	&3 \\
        SparkSQL	&7	&1	&2	&6	&0	&43	&18	& 77    &11	&11	&0	&0	&22 \\
        TiDB	    &19	&6	&7	&5	&1	&13	&5	& 56    &2	&5	&4	&1	&12 \\      
        \bottomrule
\textbf{Avg:}&12	&5	&7	&3	&1	&13	&7	& 48        &6  &6  &10  &8  &30
    \end{tabular}
\vspace{-5mm}
\end{table*}

\subsection{Operations}
\emph{Identification.} We identified operations from the source code of all DBMSs, except for SQL Server, whose source code is not public. MongoDB, MySQL, PostgreSQL, and TiDB specify operations in enumeration variables or lists. Neo4j and SparkSQL define each operation as a class or structure. SQLite defines operations as strings that are passed to the query plan generation process. We found that only SQL Server and Neo4j provided detailed documents for operations, while the other DBMSs' documents had incomplete lists of operations, and relied on illustrative examples.

\emph{Classification.} We classified operations into seven categories, as shown in the left part of \autoref{tab:number}. 
Our classification is based on the well-established relational algebra theory~\cite{codd1970relational}, which uses algebraic structures for modeling data and defining queries on it. Relational algebra operates on homogeneous sets of tuples $S=\{(s_{j1}, s_{j2},..., s_{jn}|j \in 1...m)\}$, where we commonly interpret $m$ to be the number of tuples in a table and $n$ to be the number of attributes in a tuple. We considered basic operations: selection $\sigma$, projection $\Pi$, join $\bowtie$, cartesian product $\times$, set operations $\cup$, $\cap$, $-$, aggregation $\gamma$, and rename $\rho$. Although more specific join algebra operators are available, such as left join: $\leftouterjoin$, we used $\bowtie$ to represent all kinds of join operations for simplicity. DBMSs define specific operations in query plans to realize the above algebra except for $\rho$.
On average, every DBMS defines 48 operations in query plans. Neo4j has the most operations, while InfluxDB has no operations. We noticed that Neo4j requires more operations on nodes and relationships of the graph data model, such as the operation to set node properties~\cite{neo4jset}, and SQLite is a lightweight DBMS with a limited number of operations, such as lacking the operations for creating tables. Overall, NoSQL DBMSs have more operations than relational DBMSs. An exception is InfluxDB, which does not define operations. InfluxDB's operations are disregarded in query plans due to the limited set of operations supported by the single-tuple time-series data.

\emph{Producer.} The \emph{Producer} category consists of the operations that retrieve data from storage or return constants instead of from children's operations. 
These operations realize the algebra operator $\sigma$, which selects data that satisfies a given predicate.
The operations in the \emph{Producer} category are data sources of queries, so they are typically leaf nodes of query plans. For example, in \autoref{lst:example}, the operation \textbf{\footnotesize SEARCH} in line 43 represents a full table scan, and \textbf{\footnotesize Bitmap Heap Scan} in line 28 represents a data scan from bitmaps in heap memory. Six of nine DBMSs define more than ten operations in the \emph{Producer} category, because reading data is usually expensive, and thus reads are customized for different scenarios aiming to improve efficiency. For example, indexes~\cite{guttman1984r, lehman1985study} can be used to efficiently read data.

\emph{Combinator.} The \emph{Combinator} category consists of the operations that change the permutation and combination of tuples, such as sort and union, with no changes to attributes. 
These operations realize the algebra operators $\cup$, $\cap$, and $-$.
In \autoref{lst:example}, the operation \textbf{\footnotesize Append} of PostgreSQL merge tuples from different children operations to a single set by the operations in lines 13 and 28, and is associated with query clause \lstinline{UNION} in line 8. To execute a similar functionality in SQLite, the operations \textbf{\footnotesize COMPOUND QUERY} and \textbf{\footnotesize UNION} combine data objects by the operations in lines 37 and 42, and both operations also belong to \emph{Combinator} category.

\begin{figure}
    \centering\scriptsize
    \begin{tabular}{@{}l@{}ll@{}}
        Planner COST \\
        Runtime version 5.10 \\
        \toprule
        \textbf{Operator} & \textbf{Rows} & \textbf{...} \\
        \midrule
        +ProduceResults                          & 8 & ... \\\hline
        +UndirectedRelationshipIndexContainsScan & 8 & ... \\
        \bottomrule
        Total database accesses: 5, total allocated memory: 184
    \end{tabular}
    \caption{An example query plan of the Neo4j operations of the Join category.}
    \label{fig:neo4jjoin}
\vspace{-5mm}
\end{figure}

\emph{Join.} The \emph{Join} category consists of the operations that generate new tuples by recombining attributes.
These operations realize the algebra operators $\bowtie$ and $\times$.
In \autoref{lst:example}, the operation \textbf{\footnotesize Set Join} in line 19 combines two ordered data from the operations in lines 21 and 25 based on the common fields \lstinline{t0.c0} and \lstinline{t1.c0}. MongoDB has no \emph{Join} operations, because it includes only a single document tuple for querying and lacks support for combining data from multiple documents. The higher number of \emph{Join} operations in Neo4j is because we classified the operations on the edges of the graph data model as belonging to the \emph{Join} category. In the graph data model, edges establish relationships between nodes, and a broader range of operations can be performed on the edges. For example, in Neo4j, executing the simple query \lstinline{MATCH ()-[r]->()} \lstinline{WHERE r.title ENDS WITH 'developer'} \lstinline{RETURN r} retrieves the relationships whose properties satisfy \lstinline{r.title ENDS WITH 'developer'}. The corresponding query plan is shown in \autoref{fig:neo4jjoin}. Each line in the table represents an operation and associated properties, and the content outside the table is plan-associated properties. The query plan specifies scanning the relationships, which indicates both nodes (\ie, tuples) each relationship connects, so the operation \textbf{\footnotesize UndirectedRelationshipIndexContainsScan} belongs to \emph{Join}.

\emph{Folder.} The \emph{Folder} category consists of the operations that derive new tuples from a set of tuples. 
These operations realize the algebra operator $\gamma$.
In \autoref{lst:example}, the operations \textbf{\footnotesize HashAggregate} and \textbf{\footnotesize Group} are in the \emph{Folder} category and represent data aggregation and grouping, respectively. SQLite does not define operations in the \emph{Folder} category, but shows similar information in properties together with the operations in the \emph{Producer} category. The operations in the \emph{Folder} denote DBMSs' data transformation capability, so most DBMSs support operations in the \emph{Folder} category.

\emph{Projector.} The \emph{Projector} category consists of the operations that remove attributes from all tuples. 
These operations realize the algebra operator $\Pi$.
No operation in \autoref{lst:example} belongs to this category, and 6 of 9 DBMSs have no operations in this category. We observed that these operations correspond to column lists in \lstinline{SELECT} statements, and they are not explicitly denoted in query plans.

\emph{Executor.} The \emph{Executor} category consists of the operations that make no change to tuples and attributes. 
These operations are typically DBMS-specific internal operations, and do not have a clear correspondence with any algebra operators.
In \autoref{lst:example}, the operation \textbf{\footnotesize Gather Set} in line 15 merges the data from the operation \textbf{\footnotesize Group} running in other processes for parallel execution. DBMSs define these operations to cater to various designs and goals. 
For example, PostgreSQL defines the operation \textbf{\footnotesize MEMORIZE} to cache the output from node children into memory to speed up processing. We classified these operations into the \emph{Executor} category.
The average number of operations in the \emph{Executor} category is 13, and SparkSQL has significantly more operations, 43, in the \emph{Executor} category than others, because it defines multiple operations to interact with other components, such as the Python library \emph{pandas}. 

\emph{Distributed DBMSs.}
Within our studied DBMSs, MongoDB, SparkSQL, and TiDB are distributed DBMSs, which allow executing query plans in parallel across multiple computing nodes~\cite{grove2022how}. Although they have significantly different execution steps than single-node DBMSs, their query plans have no structural difference from other DBMSs' query plans. The operations that can be executed in parallel are not explicitly stated in query plans. A major difference is that distributed DBMSs define operations to exchange data across nodes. For example, TiDB defines the operations \textbf{\footnotesize ExchangeReceiver}, \textbf{\footnotesize ExchangeSender}, and \textbf{\footnotesize Shuffle} to send, receive, and shuffle data across nodes. We classified them into \emph{Executor}. InfluxDB and Neo4j only support the distributed architecture in enterprise editions, while we studied their community editions.

\emph{Consumer.} The \emph{Consumer} category consists of operations that have no output. 
These operations correspond to non-query SQL statements, such as \lstinline{UPDATE}, so they lack a counterpart in relational algebra.
Apart from queries, which, in SQL, are \lstinline{SELECT} statements, DBMSs also support other statements, such as \lstinline{CREATE} and \lstinline{UPDATE} in SQL. DBMSs also expose query plans for these statements, and name them as execution plans for wider usage~\cite{executionplan}. The operations of the \emph{Consumer} category usually modify stored data or system variables. For example, SparkSQL uses the operation \textbf{\footnotesize SetCatalogAndNamespace} to control a particular system variable, and we assign these operations to the \emph{Consumer} category. 

\subsection{Properties}
\emph{Identification.} Each property is associated with either an operation or a query plan, and the available properties are statically encoded as strings near the generation processes of associated operations or query plans in the source code. InfluxDB's query plan representation includes only a list of plan-associated properties, while other DBMSs include both plan-associated and operation-associated properties. In \autoref{lst:example}, PostgreSQL's operations have various general properties enclosed in brackets, along with operation-specific properties in the subsequent lines. At the bottom of the serialized query plan, the property \lstinline{Planning Time} is plan-associated and represents the time to produce the query plan. In the documentation, similar to operations, only SparkSQL provides a comprehensive list of properties, while other DBMSs only show examples of properties. We explain that it is difficult to maintain the documentation of properties, which are diverse and evolving over versions. To maintain the information of properties, some third-party tools, like pgMustard, maintain a curated list of properties with accompanying explanations, but are usually commercial.

\emph{Classification.} We identified four categories of properties, as shown in the right part of \autoref{tab:number}. On average, every DBMS defines 30 properties. PostgreSQL has the most properties, since it includes many fine-grained properties. For example, it defines three properties to show the status of parallel computations: \lstinline{worker number}, \lstinline{worker launched}, and \lstinline{worker planned}, while other DBMSs provide at most one property for parallel computation. 
We give a detailed explanation of each property category as follows.

\emph{Cardinality.} The \emph{Cardinality} category consists of the numeric properties that denote the estimated data size returned by operations. The properties in this category can be associated with operations of any category or the serialized query plan as a whole. In \autoref{lst:example}, the properties \lstinline{rows} and \lstinline{width} belong to the \emph{Cardinality} category and represent the estimated number of returned rows and width. These estimates are derived from statistical information~\cite{ioannidis2003history} that DBMSs collect, such as the total number of rows and maximum values. Query plans with lower estimated cardinalities are more likely to be selected for execution during cost-based query optimization. Some DBMSs, such as MySQL, provide more fine-grained information about the number of rows that are read and returned. As a lightweight DBMS, SQLite uses simple heuristics to estimate cardinalities, and omits properties in the \emph{Cardinality} category.

\emph{Cost.} The \emph{Cost} category consists of the numeric properties that denote the estimated resource consumption. The properties in this category can be associated with operations of any category or the serialized query plan as a whole. 
In \autoref{lst:example}, the property \lstinline{cost} is in the \emph{Cost} category, and the two numbers of \lstinline{cost} denote the cost scores of starting and finishing the associated operation by estimating the total consumption of disk and CPU. As in the \emph{Cardinality} category, SQLite lacks properties in the \emph{Cost} category.

\emph{Configuration.} The \emph{Configuration} category consists of the properties that configure the operations' parameters, and their values are configuration options which are usually strings or boolean values. The properties in the \emph{Configuration} category can be associated with operations in any category or the serialized query plan, and are typically specific to operations. In \autoref{lst:example}, PostgreSQL's properties \lstinline{Group Key}, \lstinline{Set Cond}, \lstinline{Sort Key}, \lstinline{Recheck Cond}, \lstinline{Index Cond}, \lstinline{Filter} are in the \emph{Configuration} category and are specific to the associated operations to show the keys used to group, the condition to join, the key to sort, the condition to check, the index condition, and the predicate to exclude data, respectively. SQLite's property \lstinline{USING COVERING INDEX} denotes the index condition.

\emph{Status.} The \emph{Status} category consists of the properties of run-time status, and their values are runtime metrics which are usually strings or numbers. These properties can be associated with operations in any category or the serialized query plans, and typically differ depending on the operations they are attached to. In \autoref{lst:example}, the property \lstinline{Workers Planned} is in the \emph{Status} category and shows the number of available computing nodes to execute the associated operation. As another example, due to the distributed architecture, TiDB defines the property \lstinline{taskType} to show the nodes that the operation is assigned to execute on. The properties in the \emph{Status} category show running status, and are determined by the execution environment, while the properties in the Parameters are usually decided by queries. The properties in both \emph{Status} and \emph{Configuration} categories are customized, and thus are typically different across DBMSs, while the properties in other categories share similar semantics or functionalities across DBMSs.

\subsection{Formats}

\begin{table}
    \caption{The officially supported formats of query plans.}
    \label{tab:structures}
    \centering\scriptsize
    \begin{tabular}{@{}l@{}cccccc@{}}
        \toprule
                      & \multicolumn{3}{c}{\textbf{Natural}} & \multicolumn{3}{c}{\textbf{Structured}} \\
        \cmidrule(lr){2-4} \cmidrule(lr){5-7} 
        \textbf{DBMS} & \textbf{Graph}   & \textbf{Text}    & \textbf{Table} & \textbf{JSON} & \textbf{XML} & \textbf{YAML} \\
        \midrule
        InfluxDB      &                  & \checkmark       &                  &               &              &                    \\
        MongoDB       & \checkmark       &                  &                  & \checkmark    &              &                    \\
        MySQL         & \checkmark       &                  & \checkmark       & \checkmark    &              &                   \\ 
        Neo4j         & \checkmark       &                  & \checkmark       & \checkmark    &              &                    \\
        PostgreSQL    &                  & \checkmark       & \checkmark       & \checkmark    & \checkmark   &   \checkmark      \\ 
        SQL Server    & \checkmark       & \checkmark       & \checkmark       &               & \checkmark   &                  \\ 
        SQLite        &                  & \checkmark       &                  &               &              &                    \\
        SparkSQL      & \checkmark       & \checkmark       &                  &               &              &                    \\
        TiDB          & \checkmark       &                  & \checkmark       & \checkmark    &              &                    \\
        \bottomrule
    \end{tabular}
\vspace{-5mm}
\end{table}

DBMSs serialize query plans to various formats for different purposes. The formats are typically controlled by a specific configuration in queries, such as for PostgreSQL, the statement \lstinline{EXPLAIN (FORMAT JSON)} \lstinline{SELECT}\ldots serializes the query plan representation to \textsc{JSON} format. We classified all formats into two categories: \emph{natural} formats, which are optimized for readability, and \emph{structured} formats, which are optimized for machine reading. We also consider the graph formats that are supported in official IDEs.

\Autoref{tab:structures} shows the different formats of query plan representations. Overall, DBMSs support more formats in the \emph{natural} category rather than the \emph{structured} category, suggesting that DBMSs prioritize readability over machine processing. Due to the lack of a standard, none of the formats is supported by all DBMSs. For the same DBMS, the formats in the \emph{natural} category usually include less information than the formats in the \emph{structured} category. For example, in \autoref{lst:example}, the property \lstinline{Parent Relationship} represents how the associated operation passes data to another operation. This property is ignored in the text format of the \emph{natural} category, but is shown in the \textsc{JSON} format of the \emph{structured} category. We provide more details of each format as follows.

\emph{Natural category.} The \emph{natural} category includes graph, text, and table formats. Query plans are usually serialized as graphs for DBMSs' IDEs, such as Workbench\footnote{\url{https://dev.mysql.com/doc/workbench/en/wb-performance-explain.html}} for MySQL, Compass\footnote{\url{https://www.mongodb.com/docs/compass/current/query-plan/}} for MongoDB. Text formats represent query plans as plain text, such as shown in \autoref{lst:example}. Table formats encode each operation and associated properties in a line, and use line numbers to represent the tree structure of query plans. Graph formats are intuitive to understand, so graph formats are supported by most DBMSs.

\emph{Structured category.} The \emph{structured} category includes the \textsc{JSON}, \textsc{XML}, and \textsc{YAML} formats. These formats are standardized and widely used for exchanging data~\cite{ecma2013ecma, bosak1997xml}.  \textsc{JSON} is more widely supported by DBMSs than other \emph{structured} formats, and PostgreSQL supports all \emph{structured} formats. \emph{Structured} formats are not supported by some DBMSs, such as InfluxDB, SQLite, and SparkSQL. 
SQLite can output a \emph{structured} format of bytecode, which consists of low-level instructions, not operations and properties, so we do not consider it as a \emph{structured} query plan representation.

\begin{table}[]
    \caption{Third-party visualization tools for query plans.}
    \label{tab:tool}
    \centering
    \begin{tabular}{@{}lp{1.6cm}l@{}}
    \toprule
    \textbf{Tool} & \textbf{DBMSs} & \textbf{License} \\
    \midrule
Postgres Explain Visualizer 2~\cite{pev} & PostgreSQL & Open-source \\
pgmustard~\cite{pgmustard}                 & PostgreSQL & Commercial \\
pganalyze~\cite{pganalyze}                 & PostgreSQL & Commercial \\
ApexSQL~\cite{apexsql}                     & SQL Server & Commercial \\
Plan Explorer~\cite{planexplorer}          & SQL Server & Commercial \\
Azure Data Studio~\cite{azuredatastudio}   & SQL Server & Commercial \\
Dbvisualizer~\cite{dbvis} & MySQL, PostgreSQL, SQL Server & Commercial \\
    \bottomrule
    \end{tabular}
\vspace{-4mm}
\end{table}

\emph{Visualization.} Apart from the graph formats in official IDEs, third-party visualization tools show query plans based on \emph{structured} formats to enhance the readability of query plans. \autoref{tab:tool} shows the visualization tools we found for the studied DBMSs. Six of the seven tools are commercial, suggesting the value of understanding query plan representations for developers. Building these tools requires non-trivial effort, because a tool is specific to a DBMS.
\section{Unified Query Plan Representation}\label{sec:format}

Our study in \Autoref{sec:study} shows that query plan representations share the same conceptual basis, which is why we propose a unified query plan representation that is:
\begin{enumerate}
    \item complete, to include all information of a query plan,
    \item general, to support various DBMSs we studied, and,
    \item extensible, to support DBMSs we did not study.
\end{enumerate}

\begin{figure}
\begin{lstlisting}[caption={The unified query plan representation in EBNF.},captionpos=t, language=bash, label=lst:grammar, escapeinside=@@, morekeywords={plan, tree, node, operation, properties, property, operation_category, operation_identifier, property_category, property_identifier, keyword, value, string, number, boolean, capital_letter, letter, digit}]
plan ::= ( tree )? properties
tree ::= node ( '--children-->' '{' tree (',' tree)* '}' )?
node ::= operation properties
operation ::= 'Operation' ':' operation_category '->' operation_identifier
properties ::= ( property ( ',' property )* )?
property ::= property_category '->' property_identifier ':' value
operation_category ::= 'Producer' | 'Combinator' | 'Join' | 'Folder' | 'Executor' | 'Projector' | 'Consumer'
property_category ::= 'Cardinality' | 'Cost' | 'Configuration' | 'Status'
operation_identifier ::= keyword
property_identifier ::= keyword
keyword ::= letter ( letter | digit | '_' )*
value ::= string | number | boolean | 'null'
string ::= '"' ( letter | digit )* '"'
number ::= '-'? digit+
boolean ::= 'true' | 'false'
letter ::= [a-zA-Z]
digit ::= [0-9]
\end{lstlisting}
\vspace{-5mm}
\end{figure}

\subsection{Design}
To define and illustrate the unified query plan representation, we adopted Extend Backus Naur Form (EBNF)~\cite{feynman2016ebnf}, which is a metasyntax notation to express context-free grammars. \Autoref{lst:grammar} shows the unified query plan representation in EBNF. Following our study in \Autoref{sec:study}, the unified representation includes the three identified conceptual components of different categories. We define \textbf{\small plan} as a \textbf{\small tree} that can have plan-associated \textbf{\small properties}. Within the \textbf{\small tree}, a \textbf{\small node} is defined as an \textbf{\small operation} and zero or multiple operation-associated \textbf{\small properties}. \textbf{\small operations} and \textbf{\small properties} are key-value pairs including corresponding \textbf{\small categories}, \textbf{\small identifiers}, and \textbf{\small values}. 

We use a unified naming convention to denote operations and properties in the unified query plan representation. A unified naming convention increases the readability and consistency of the representation while avoiding name collisions. \Autoref{sec:study} shows that various operations and properties share similar semantics, so we mapped DBMS-specific names of operations and properties to unified names. For example, we mapped the operation name \textbf{\footnotesize Seq Scan} in PostgreSQL, \textbf{\footnotesize Table Scan} in SQL Server, and \textbf{\footnotesize TableFullScan} in TiDB to \textbf{\footnotesize Full Table Scan}.

\subsection{Analysis}
We analyze whether this design achieves the three goals.

\emph{Completeness.} The unified query plan representation includes the three conceptual components: operations, properties, and formats, which we identified in \Autoref{sec:study}. Operations and properties are included in the tree of the unified representation, and the unified representation can be serialized into other standard formats, such as \textsc{JSON} and \textsc{XML}, which are used in query plan representations.

\emph{Generality.} The unified query plan representation supports the query plan representations of the nine DBMSs that we studied. InfluxDB's query plan includes a list of properties without operations, which can be represented by plan-associated properties in the unified representation. For the other DBMSs, we identified operations and properties, mapped them into unified names, and organized them into our unified representation.

\emph{Extensibility.} The definitions of operations, properties, and categories in the unified query plan representation can be extended or shrunk while keeping forward and backward compatibility. Forward compatibility refers to allowing a system to accept input intended for a later version of itself, and backward compatibility refers to allowing a system to accept input intended for an older version of itself~\cite{zeldman2003designing}. Both evaluate whether the applications based on the unified query plan representation still work if we update the representation to support more or different DBMSs. To keep forward compatibility, we can add more categories by expanding \textbf{\small operation\_category} to include more category names, and add more operations and properties whose names comply with the definition of the keyword at line 11 at \Autoref{lst:grammar}.
The required effort is minimal, as it involves only adding keyword definitions for any new operation or property in a specific DBMS, while the rest of the unified query plan representation remains unaffected.
An existing application still can parse the revised representation by ignoring the newly added categories, operations, and properties, or handling them in a generic way (\eg, a visualization tool could represent unknown operations using a generic visual shape). For backward compatibility, similarly, an existing application still can parse the old version of the representation, whose categories, operations, and properties are included in the new version of the representation, so no maintenance effort is required. 

\emph{An Example of Extensibility.}
We assume that one of the DBMSs, PostgreSQL,
introduces a new operation in query plans to implement Large Language Model (LLM)-based joining~\cite{urban2024efficient}.
The \method developers would require minimal effort to accommodate this feature by adding the keyword \textbf{\small LLM Join} for the new operation, without impacting the rest of the unified query plan representation. 
Similarly, to deprecate this feature, we would remove the keyword in the grammar.
These changes would be both forward and backward-compatible to the applications of \method.
Suppose a visualization tool supports \method to visualize query plans for our studied DBMSs.
It is forward-compatible as \method provides the information of \textbf{\small LLM Join} and the visualization tool can visualize it without any modification.
It is also backward-compatible as older versions of \method would still be supported by the visualization tool, as any older formats would have at most as many operations as the currently-supported format.
\section{Applications}
We implemented the prototype of a reusable library, \method, to maintain the unified query plan representation. While our main motivation is facilitating the implementation of automated testing approaches, the unified representation also enables other applications of visualization and benchmarking:

\begin{description}
    \item[\textbf{A.1 Testing.}] The DBMS testing methods \qpg and \cert were implemented in a DBMS-specific way due to DBMS-specific query plan representations, and we show how \method allows both methods to be implemented in a DBMS-agnostic way.
    \item[\textbf{A.2 Visualization.}] Visualization tools visually display serialized query plans to ease understanding, but are typically specific to a particular DBMS. We show a general visualization tool based on \method.
    \item[\textbf{A.3 Benchmarking.}] Benchmarking is an important method to evaluate the performance of DBMSs. We show a case analysis of a comparison of query plan representations using \method. We hope that allowing developers to easily compare different DBMSs' query plans enables them to improve their DBMSs' query optimization capabilities.
\end{description}

\emph{DBMSs.} For A.2 and A.3, we used MongoDB, MySQL, Neo4j, PostgreSQL, and TiDB, because they support the \textsc{JSON} format of query plans. \textsc{JSON} is the most widely supported structural format, and it typically includes more detailed information than other formats. For example, JSON may include read and optimization costs, but the text format often only includes overall cost, as it is typically optimized for human readability. Within these DBMSs, we used MySQL, PostgreSQL, and TiDB for A.1 as they are supported by \sqlancer. We used the same versions of DBMSs that we studied in the \autoref{tab:benchmark}.

\emph{Data set.} For A.1, we used \sqlancer to generate test cases. For A.2 and A.3, we used the benchmarks TPC-H~\cite{tpch} for the five DBMSs, YCSB~\cite{cooper2010benchmarking} for MongoDB, and WDBench~\cite{angles2022wdbench} for Neo4j.
TPC-H is one of the most established benchmarks consisting of business-oriented ad-hoc queries and concurrent data modification for evaluating relational DBMSs. It comprises 8 tables and 22 queries for relational DBMSs: MySQL, PostgreSQL, and TiDB in our experiments.
For MongoDB and Neo4j, TPC-H requires manual effort to execute as it was not designed for non-SQL languages and non-relational data models. MongoDB is based on a document data model, which lacks support for join operations, so we embedded all entities in one document and rewrote queries 1, 3, and 4 in the MongoDB Query Language (MQL), following a tutorial~\cite{mongodbtpch}. For Neo4j, which is based on a graph data model, we mapped nodes into rows of the relational data model and edges into foreign keys of the relational data model following another tutorial~\cite{neo4jtpch}, and rewrote queries 1--14, 16--19 using the Cypher Query Language (CQL) following an example~\cite{cqltpch}.
WDBench evaluates graph DBMSs and consists of 2623 queries. YCSB is used for NoSQL DBMSs and dynamically generates workloads including updates and queries. We used both benchmarks to further explore the query plans in MongoDB and Neo4j.

\emph{Implementation.} \method is a reusable library that consists of around 300 lines of Python code to implement a reusable library, which allows adding or updating operations and properties. The prototype supports serializing query plan representations into \textsc{JSON} as well as text formats, and provides an interface for supporting more formats. We also implemented five customized converters to parse the query plan representations from existing query plan representations to the unified query plan representation, and each parser has around 200 lines of code.

\subsection*{A.1 Testing}
We show how \method allows implementing the testing methods \qpg~\cite{qpg} and \cert~\cite{cert} in a DBMS-agnostic way.
\qpg is a test case generation approach that is guided by query plans; specifically, it mutates a database if no new query plans have been observed for a specific number of randomly generated queries, aiming to subsequently exercise new query plans, and thus exploring ``interesting'' behaviors. In terms of implementation, evaluating whether a query plan is structurally different from another requires ignoring unstable information, such as random identifiers and the estimated cost in query plans. \cert is a test oracle for finding performance issues by comparing estimated cardinalities, which have to be extracted from query plans. Both methods were implemented in \sqlancer, which is a popular and widely used tool for automatically testing DBMSs. The original implementations used DBMS-specific and error-prone methods, such as string matching and substitution, which had to be implemented for every DBMS that was supported. Both approaches support relational DBMSs, but were not applied to the popular open-source DBMSs MySQL and PostgreSQL, as additional DBMS-specific parsers would have been required. Based on \method, we implemented general parsers for both methods to support all \method-compatible DBMSs. We did not consider \mozi, because its source code is not public.

\begin{figure}
    \centering
    \includegraphics[width=0.8\columnwidth]{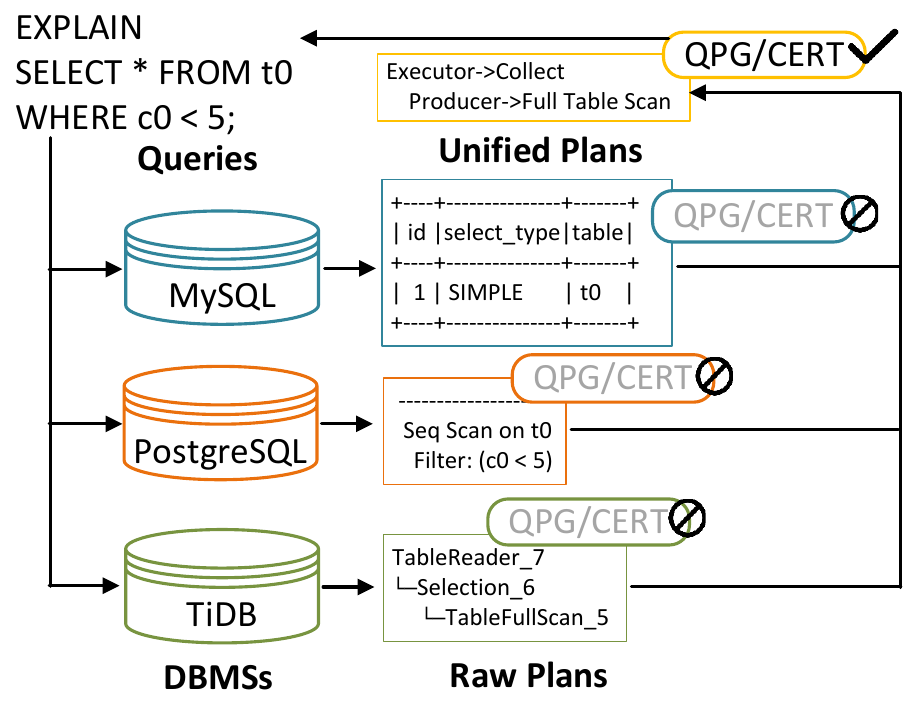}
    \caption{The architecture of using \method for \qpg and \cert. Grey font means the implementation without \method, and $\checkmark$ refers to the implementation with \method. \qpg/\cert stops when observing an unexpected result or query plan.}
    \label{fig:testing}
\vspace{-5mm}
\end{figure}

\autoref{fig:testing} shows the architecture of using \method for \qpg and \cert. Both \qpg and \cert iteratively execute queries in target DBMSs and retrieve query plans for generating the next query and examining cardinalities respectively. In this example, suppose both \qpg and \cert execute the query \lstinline{EXPLAIN SELECT * FROM t0 WHERE t0<5} in MySQL, PostgreSQL, and TiDB, the query plans returned by three DBMSs are significantly different. Without \method, \qpg and \cert require DBMS-specific implementation to parse the raw query plans. With \method, we converted raw query plans into the unified query plan, and \qpg and \cert require only a single implementation based on \method. In this example, TiDB's plan is converted into two operations, while PostgreSQL's and MySQL's plans are converted into \lstinline{Producer->Full Table Scan} only as TiDB requires \lstinline{Executor->Collect} to receive data from other nodes in a distributed system.

To evaluate \method on \qpg and \cert, we applied the general parser to MySQL, PostgreSQL, and TiDB. We ran our revised versions of \qpg and \cert for 24 hours and found \numbugs unique and previously unknown bugs, as shown in \Autoref{tab:bugs}. Developers confirmed 16 of the \numbugs bugs and fixed two bugs. The one bug in PostgreSQL was still waiting for the response from developers, so we did not report more bug reports to avoid burdening developers. 11 of \numbugs bugs are \emph{Critical}, \emph{Serious}, or \emph{Major}, which represent that the bugs can significantly affect the systems. Additionally, \cert found hundreds of potential bug-inducing test cases in 24 hours, but it is challenging to distinguish their uniqueness, which requires developers' expertise. \Autoref{lst:bug} shows a bug in MySQL found by the test case generated by \qpg with \method.
Note that we identified this bug by the test oracle Ternary Logic Partitioning (TLP)~\cite{Rigger2020TLP}; however, for presentation, we simplified the bug-inducing test case by demonstrating that the same query returns different results in lines 4 and 6 depending on whether the index exists. 
The cause of the bug was an incorrect table look-up due to the index inserted by the SQL statement in line 5. Using \method, we were able to apply \qpg to MySQL easily, which was previously incompatible and untested by \qpg. This enabled us to find this bug in MySQL.

\begin{table}
    \centering
    \caption{Previously unknown and unique bugs found by \qpg with \method.}
    \begin{tabular}{@{}lllll@{}}
        \toprule
        \textbf{DBMS} & \textbf{Found by} & \textbf{Bug ID}  & \textbf{Status} & \textbf{Severity} \\
        \midrule
MySQL	  & \qpg    & \href{https://bugs.mysql.com/bug.php?id=113302}{113302} &  Confirmed & Critical      \\
MySQL	  & \qpg    & \href{https://bugs.mysql.com/bug.php?id=113304}{113304} &  Confirmed & Critical      \\
MySQL	  & \qpg    & \href{https://bugs.mysql.com/bug.php?id=113317}{113317} &  Confirmed & Critical      \\
MySQL	  & \qpg    & \href{https://bugs.mysql.com/bug.php?id=114204}{114204} &  Confirmed & Serious       \\
MySQL	  & \qpg    & \href{https://bugs.mysql.com/bug.php?id=114217}{114217} &  Confirmed & Serious       \\
MySQL	  & \qpg    & \href{https://bugs.mysql.com/bug.php?id=114218}{114218} &  Confirmed & Serious       \\
MySQL	  & \cert   & \href{https://bugs.mysql.com/bug.php?id=114237}{114237} &  Confirmed & Performance   \\
PostgreSQL& \cert   & \href{https://postgrespro.com/list/id/SG2PR06MB2810355511BEA44B808B32178A4B9@SG2PR06MB2810.apcprd06.prod.outlook.com}{Email} &  Pending   & Performance   \\
TiDB	  & \qpg    & \href{https://github.com/pingcap/tidb/issues/49107}{49107} &  Fixed     & Major         \\
TiDB	  & \qpg    & \href{https://github.com/pingcap/tidb/issues/49108}{49108} &  Confirmed & Major         \\
TiDB	  & \qpg    & \href{https://github.com/pingcap/tidb/issues/49109}{49109} &  Fixed     & Major         \\
TiDB	  & \qpg    & \href{https://github.com/pingcap/tidb/issues/49110}{49110} &  Confirmed & Major         \\
TiDB	  & \qpg    & \href{https://github.com/pingcap/tidb/issues/49131}{49131} &  Confirmed & Major         \\
TiDB	  & \qpg    & \href{https://github.com/pingcap/tidb/issues/51490}{51490} &  Confirmed & Moderate      \\
TiDB	  & \qpg    & \href{https://github.com/pingcap/tidb/issues/51523}{51523} &  Confirmed & Moderate      \\
TiDB	  & \cert   & \href{https://github.com/pingcap/tidb/issues/51524}{51524} &  Confirmed & Minor         \\
TiDB	  & \cert   & \href{https://github.com/pingcap/tidb/issues/51525}{51525} &  Confirmed & Minor         \\
        \bottomrule
    \end{tabular}
    \label{tab:bugs}
\vspace{-5mm}
\end{table}

\begin{figure}
\begin{lstlisting}[caption={A bug found by \qpg with \method.},captionpos=t, label=lst:bug, escapeinside=&&]
CREATE TABLE t0(c0 INT, c1 INT);
INSERT INTO t0(c1, c0) VALUES(0, 1);

SELECT * FROM t0 WHERE t0.c1 IN (GREATEST(0.1, 0.2)); -- empty result &\oksymbol&
CREATE INDEX i0 ON t0(c1);
SELECT * FROM t0 WHERE t0.c1 IN (GREATEST(0.1, 0.2)); -- {1|0} &\bugsymbol&
\end{lstlisting}
\vspace{-7mm}
\end{figure}

We also found that \method reduces the risk of implementation bugs for DBMS-specific query plan parsers. We identified an implementation bug in \qpg. Specifically, the query plan parser for TiDB failed to exclude random identifiers due to an incorrect parameter for \lstinline{EXPLAIN}. With the single implementation for a parser and the unified query plan representation, we have a lower risk of introducing these implementation bugs.

\result{\method enables large-scale adoption for testing methods \qpg and \cert in a DBMS-agnostic implementation way.}

\subsection*{A.2 Visualization}

We implemented a visualization tool for serialized query plans by modifying PEV2~\cite{pev}, a customized query plan visualization tool for PostgreSQL, to use the unified query plan representation.
We modified its parser to support identifying the unified query plan representation, and updated its definitions of visualized elements, such as operation names.

\begin{figure}
    \centering
    \includegraphics[width=0.95\columnwidth]{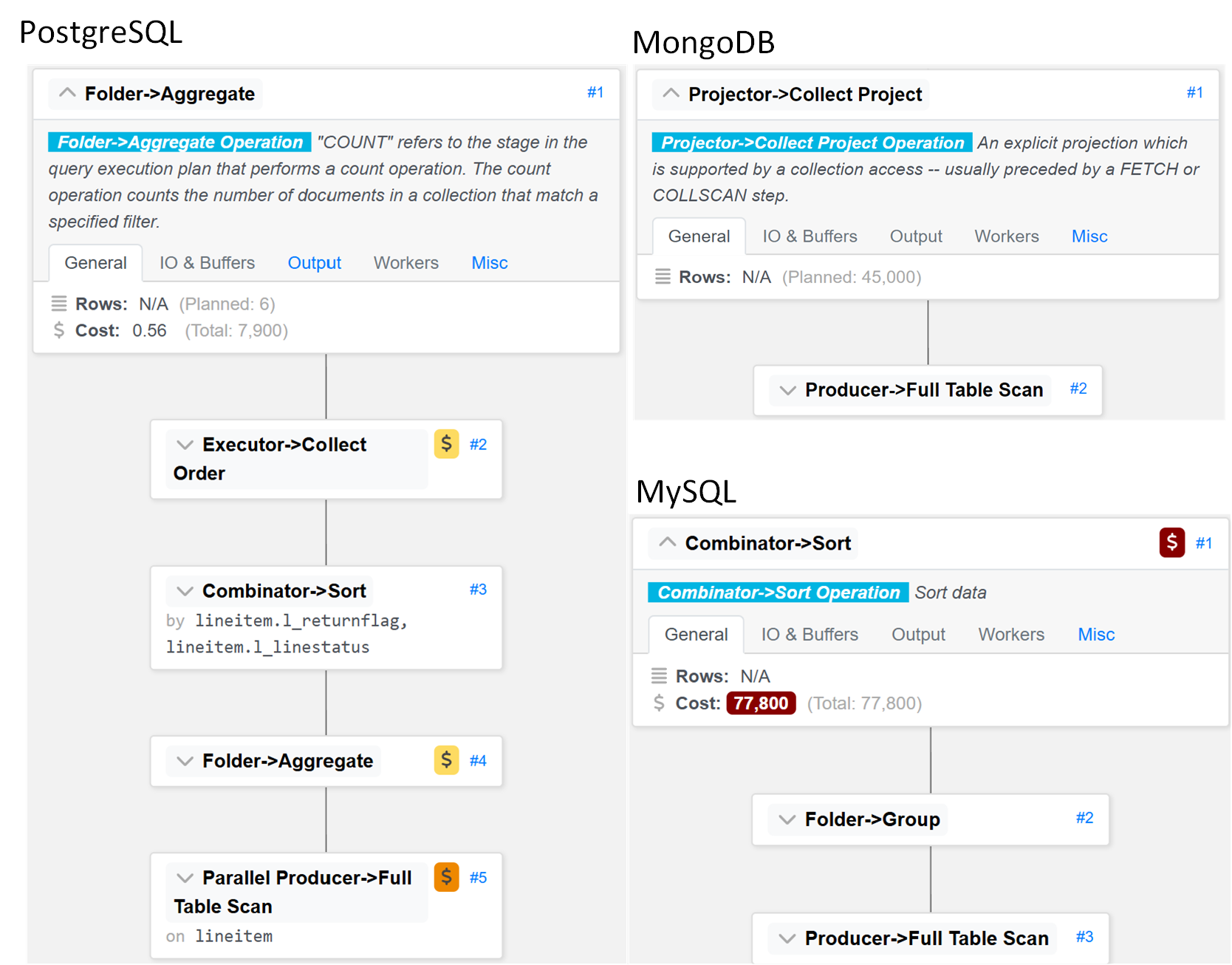}
\vspace{-3mm}
    \caption{Visualized unified query plans of query 1 from TPC-H benchmark.}
    \label{fig:visualization}
\vspace{-5mm}
\end{figure}

\method significantly reduces the effort to build visualization tools on query plans. According to the Git repository, developers of PEV2 committed 24,559 lines of code within the 188 days between the initial commit and the first release. As a naive, but reasonable approximation, we assumed this period reflects the minimal required effort to build a visualization tool. Thus, implementing separate visualization tools for the studied DBMSs would require around $188 * 5 = 940$ days. Using \method, we modified only around 800 lines of code of PEV2 to support all DBMSs. For PEV2, the average development speed is $24,559 / 188 \approx 130$ lines of code per day, so the estimated effort for adapting \method is $800 / 130 \approx 6$ days.  Compared to implementing DBMS-specific visualization tools, which would require 940 days, \method would take only $194$ days ($188 + 6$) and would reduce the time effort by approximately 80\%. Furthermore, the percentage of effort reduction would increase as the number of supported DBMSs grows. Note that this approximation ignores many factors, such as the complexity of query plans and the time to understand visualization tools.

\autoref{fig:visualization} shows examples of visualized unified query plans in PostgreSQL, MongoDB, and MySQL. These query plans stem from query 1 of the TPC-H benchmark. An operation and its associated properties is depicted as rectangle, representing a node. For example, in the first node of MySQL query plan, \textbf{\footnotesize Combinator$\rightarrow$Sort} represents the operation \textbf{\footnotesize Sort}, which belongs to the category \textbf{\footnotesize Combinator}. The following is the description and properties.
Our unified query plan representations enable a general visualization tool that can visualize query plans from any DBMS that exposes query plans.

\result{Existing DBMS-specific visualization tools could support more DBMSs if they supported our unified query plan representation.}

\subsection*{A.3 Benchmarking}
In this application, we show how to find potential optimization opportunities by comparing the serialized query plans across DBMSs using the unified query plan representation. Query optimization is a critical process for DBMSs' performance. To evaluate how effective a query optimization is, existing methods depend on measuring DBMSs' execution time on standard datasets, such as TPC-H~\cite{tpch} and the Join Order Benchmark (JOB)~\cite{leis2015good}. The execution time shows the overall performance difference between various DBMSs, but cannot provide possible reasons for performance gaps across DBMSs. \method enables comparing various serialized query plans across DBMSs. Specifically, we collected metrics on the number of operations in DBMSs' query plan representations and show an example of analyzing the difference.

\begin{table}
    \caption{Average number of operations in query plans from TPC-H.}
    \label{tab:application2}
    \centering\scriptsize
    \begin{tabular}{@{}l@{}rrrrrrr@{}}
    \toprule
    \textbf{DBMS} & \textbf{Prod.} & \textbf{Comb.} & \textbf{Join} & \textbf{Folder} & \textbf{Proj.} & \textbf{Exec.} & \textbf{Sum} \\
    \midrule
MongoDB	    & 1.00	& 0.00	& 0.00	& 0.00	& 1.00	& 0.00	& 2.00 \\
MySQL	    & 4.55	& 0.82	& 2.77	& 0.86	& 0.00	& 0.27	& 9.27 \\
Neo4j	    & 0.39	& 0.78	& 2.89	& 0.06	& 0.72	& 3.06	& 7.89 \\
PostgreSQL	& 3.95	& 1.32	& 2.64	& 1.73	& 0.00	& 2.45	& 12.09 \\
TiDB	    & 4.18	& 0.82	& 2.73	& 1.41	& 1.77	& 3.73	& 14.64 \\
    \bottomrule
    \end{tabular}
\vspace{-3mm}
\end{table}

\begin{table}
    \caption{Average number of operations in query plans from YCSB for MongoDB and WDBench for Neo4j.}
    \label{tab:application2revision}
    \centering\scriptsize
    \begin{tabular}{@{}l@{}rrrrrrr@{}}
    \toprule
    \textbf{DBMS} & \textbf{Prod.} & \textbf{Comb.} & \textbf{Join} & \textbf{Folder} & \textbf{Proj.} & \textbf{Exec.} & \textbf{Sum} \\
    \midrule
MongoDB	    & 1.00	& 0.00	& 0.00	& 0.00	& 0.00	& 0.00	& 1.00 \\
Neo4j	    & 0.45	& 0.00	& 3.73	& 0.00	& 0.37	& 1.44	& 5.99 \\
    \bottomrule
    \end{tabular}
\vspace{-3mm}
\end{table}

\autoref{tab:application2} shows the average number of operations in each category for the query plans of queries from the TPC-H benchmark. We omitted the \emph{Consumer} category as we did not encounter any such operations. Overall, the relational DBMSs, MySQL, PostgreSQL, and TiDB, have more operations than the non-relational DBMSs, MongoDB and Neo4j. It is because relational DBMSs have more operations in the \emph{Producer} category. To further explain the difference in the \emph{Producer} category, we looked into query plans and found that each table in relational DBMSs requires at least one operation to read data, while non-relational DBMSs usually read all data in one or two operations. Apart from MongoDB, the other four DBMSs have a similar number of \emph{Join} operations. The results also show that the query plans of the queries in the TPC-H benchmark suite usually do not cover all categories of operations due to the limited set of queries. 

Similarly, \autoref{tab:application2revision} shows the average number of operations for the query plans from the YCSB and WDBench benchmarks respectively. Both benchmarks are designed specifically for NoSQL and graph DBMSs, but mainly consider input diversity instead of internal execution diversity, as query plans are DBMS-specific and challenging to examine. As a result, compared to TPC-H, YCSB and WDBench have a similar distribution of operations. For some categories, the query plans in YCSB and WDBench include even fewer operations than those in TPC-H. For example, YCSB does not expose any operation in the \emph{Projection} category and WDBench does not expose any operation in the \emph{Combinator} and \emph{Folder} categories. This shows that the TPC-H benchmark can efficiently expose operations in query plans.

\begin{figure}
    \centering
    \includegraphics[width=\columnwidth]{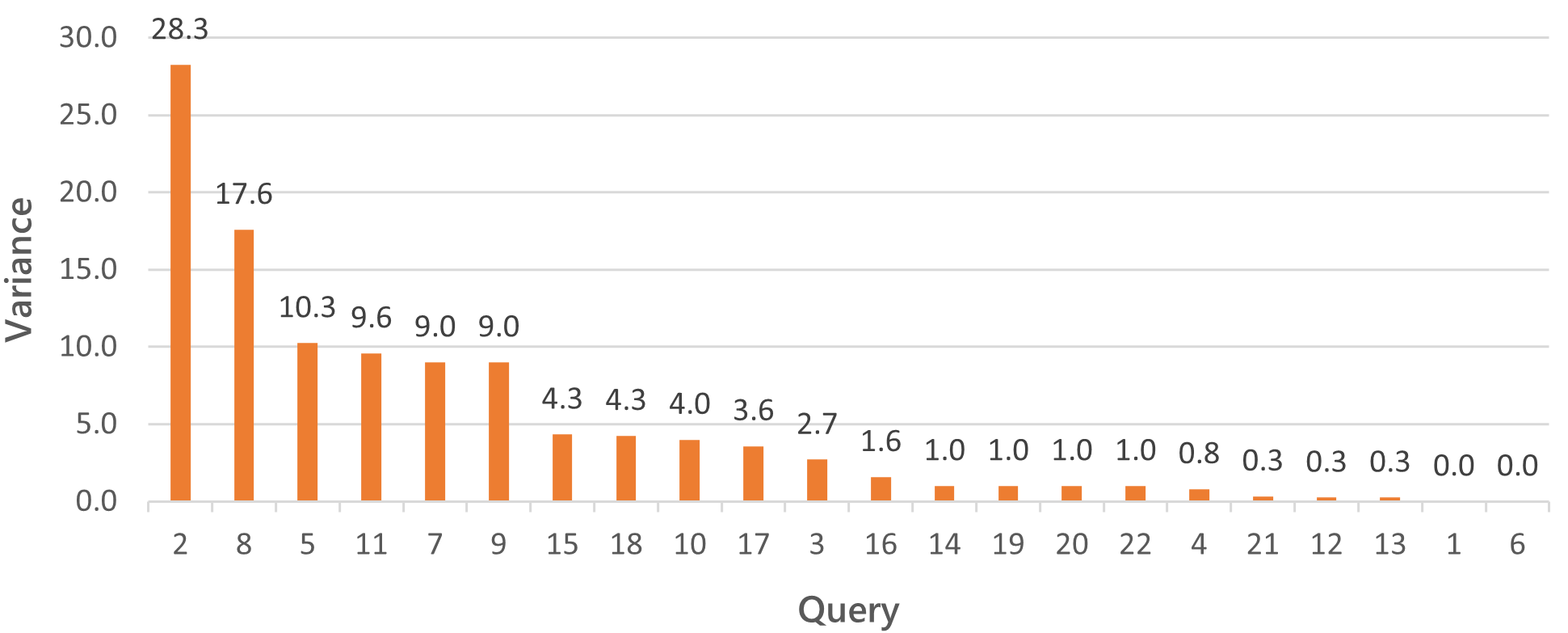}
    \caption{Variance of the number of Producer operations for each query in TPC-H benchmark across five DBMSs.}
    \label{fig:variance}
\vspace{-5mm}
\end{figure}

To find potential optimization opportunities, we analyzed the operations of the \emph{Producer}. Across the five DBMSs, the operations of the \emph{Producer} category are typically expensive for performance, and hence, developers typically aim to analyze and reduce the occurrence of the operations in this category.
\autoref{fig:variance} shows the variance of the number of operations in the \emph{Producer} category for each query plan. Among 22 queries, the variances of six queries are more than 5, indicating a significant difference. Queries 2, 8, 5, 7, and 9 have a significant variance due to different data models. For example, for query 2, the DBMSs of relational data models, MySQL, TiDB, and PostgreSQL, have 10, 12, and 9 operations, while the DBMS based on the graph data model, Neo4j, has only 1 operation in the \emph{Producer} category. Query 11 has a significant variance due to a potential optimization issue, and we explain it as follows.

\begin{figure}
\begin{lstlisting}[caption={The unified serialized query plan representations in the text format for query 11 in TPC-H. Underlines represent table names. Predicates in the query and properties in the query plan representations are ignored for brevity.},captionpos=t, label=lst:application2, escapeinside=@@, morekeywords={Folder, Join, Producer, Executor, Projector, Combinator}]
SELECT ... FROM PARTSUPP, SUPPLIER, NATION WHERE ...
HAVING ... > (SELECT ... FROM PARTSUPP, SUPPLIER, NATION WHERE ...) ...;
--------------------------------------------------------
PostgreSQL:            TiDB:
Combinator->Sort       Projector->Project
 Folder->Aggregate      Combinator->Sort
  Join->Hash Join        Folder->Aggregate Hash
   Producer->Full Table   Projector->Project
   name object: @\underline{partsupp}\,@    Join->Index Hash
   Executor->Hash Row       Join->Index Hash
    Join->Hash               Executor->Collect
     Producer->Full Table     Producer->Full Table
     name object: @\underline{supplier}\,@     name object: @\underline{nation}@ 
     Executor->Hash Row      Executor->Collect Order
      Producer->Full Table    Producer->Index-only...
      name object: @\underline{nation}\,@      name object: @\underline{supplier}@ 
 Folder->Aggregate          Executor->Collect Order
  Join->Hash Join            Producer->Index-only...
   Producer->Full Table      name object: @\underline{partsupp}@ 
   name object: @\underline{partsupp}\,@       Producer->Id Scan
   Executor->Hash Row         name object: @\underline{partsupp}@ 
    Join->Hash Join
     Producer->Full Table
     name object: @\underline{supplier}@ 
     Executor->Hash Row
      Producer->Full Table
      name object: @\underline{nation}@
\end{lstlisting}
\vspace{-5mm}
\end{figure}

\autoref{lst:application2} shows query 11 from the TPC-H benchmark suite and the corresponding serialized query plans of PostgreSQL and TiDB using the unified query plan representations in text format.
The query references the three tables \lstinline{PARTSUPP, SUPPLIER, NATION} twice in the \lstinline{FROM} and \lstinline{HAVING} clauses respectively. PostgreSQL uses six table scans, one for each table reference in the original query, while TiDB could optimize the query to use only three scans. The table \lstinline{partsupp} is scanned twice in lines 19 and 21, because TiDB reduces the data size of table scanning by retrieving a secondary index before the scan operation. The first scan retrieves indexes only to obtain the row id, which is used for the second scan. Suppose both DBMSs execute the operations in order and the same operation has the same performance overhead in both DBMSs, then the query plan with three table scans is more efficient than the query plan with six table scans.

We quantitatively evaluated the potential performance improvement contributed by our analysis. It is challenging to modify PostgreSQL source code to more effectively optimize this query plan. To estimate the performance improvement, we evaluated the actual execution time for the six table scans in PostgreSQL's query plan using the SQL prefix \lstinline{EXPLAIN ANALYZE}, which executes the query and returns the execution information of the query plan including the execution time for each operation. For 10 GB data of TPC-H benchmark, the whole query consumes 1503ms, and the six table scans consume 214ms, 5ms, 1ms, 216ms, 2ms, and 189ms, respectively. If PostgreSQL optimizes this query plan to remove the last three table scans, the execution time would be significantly reduced by 407ms ($216+2+189$), which accounts for 27\% of the overall execution time of the query. Note that we assumed that eliminating the three operations would not influence the execution time of other operators.
\method enables this analysis and provides actionable insight for DBMS developers to improve query performance by reducing three repeated table scans. We reported this issue to the PostgreSQL developers, who confirmed that this case indicates an unsupported optimization by PostgreSQL. One developer considered how this could be supported---"\emph{[...] maybe there is some easy way to hook this into the same code used by GROUPING SETS [...]"}.

\result{Comparing query plans between DBMSs, using the unified query plan representation, provides insights for improving query optimizers.}

\section{Discussion}

\emph{Paths to adoption.} Developers can use the unified query plan representation by customized converters to convert original serialized query plans into the unified query plan representation. For all three applications we showed, we implemented five customized converters, each of which has around  200 lines of code, within one week only. If the converters can be implemented by DBMS developers or experts, who are well-versed in the query plans, it is plausible that a higher-quality converter could be developed in a shorter time. We hope that DBMSs will directly expose query plans in the unified representation in the long term, thus avoiding a conversion.

\emph{Additional use cases.} We envision several additional use cases that are enabled by our unified query plan representations. Toward a comprehensive evaluation of query optimization, additional metrics could be explored, such as similarity on tree structures~\cite{yang2005similarity}, to compare different DBMSs' query plans using our unified query plan representation. To find issues in query optimization, we can apply differential testing~\cite{mckeeman1998differential} or other methods to compare the unified query plan representations among DBMSs. Query optimization approaches based on machine learning have been proposed that take query plans as input and output suggestions for indexes~\cite{ding2019ai}, views~\cite{yuan2020automatic}, and join orders~\cite{marcus2018deep, yu2020reinforcement}, so our unified query plan representation would allow exchanging training data in different DBMSs to improve the performance of models.

\emph{Substrait.} \method complements Substrait~\cite{substrait}, which provides a DBMS-agnostic specification for logical query plan representation to enable cross-language serialization for relational algebra. 
First, Substrait mainly focuses on logical query plans, while \method focuses on physical query plans. Many physical operations are DBMS-specific and are not considered by Substrait. In \autoref{sec:study}, our study based on real-world query plans shows that the physical operations of \emph{Executor} and \emph{Consumer} do not have corresponding operators in relational algebra, and Substrait does not support them. Integrating \method as a backend component for Substrait would allow it to analyze, optimize, and test DBMS-specific query executions comprehensively.
Second, \method enables different application domains compared to Substrait. \method aims to reduce the effort of building applications on query plans instead of cross-language serialization for relational algebra. The unified logical query plans, provided by Substrait, cannot facilitate building general applications based on physical query plans such as demonstrated in this work.
Lastly, \method offers easier integration with real-world DBMSs. Unlike Substrait, which requires additional support for DBMSs to both output logical query plans in Substrait format and execute those plans, \method directly parses existing query plans without requiring any modifications to the DBMS. This significantly reduces the implementation effort typically needed by Substrait.

\emph{Completeness.} We believe our study and implementation are sufficiently complete as we studied query plans from various sources. As we observed in our study, DBMSs lack detailed and standard documents for query plans, so we comprehensively studied query plans from documents, source code, and concrete executions with the workloads from TPC-H, WDBench, and YCSB benchmarks. We also evaluated our implementation of \method on visualization and benchmarking with three benchmarks. 
However, the possibility of missing features in query plans still exists. For example, some properties may be dynamically generated without detailed documentation. To observe them, we require specific workloads, which we may overlooked. To alleviate this potential issue, we designed an extensible \method. Such missing properties can be easily integrated into the unified query plan representation by adding keyword definitions as we explained in \autoref{sec:format}.

\emph{Threats to validity.} 
Our study faces several threats to validity, which denotes the trustworthiness~\cite{lincoln1985naturalistic, host2012case} of the results, and to what extent they are unbiased. 
A major concern is the degree to which the data and analysis depend on the specific researchers. We followed the best practice of triangulation~\cite{host2012case}, which refers to taking multiple perspectives toward the same object, to increase reliability. For data triangulation, we collected data from multiple sources: documentation, source code, and third-party applications. For observer triangulation, one author conducted the study, and another author validated the findings against the raw data. For methodological triangulation, we used a qualitative method to analyze query plan representations, and a quantitative method to examine and classify the three conceptual components of query plan representations. Furthermore, we have made the process of data collection and analysis publicly available, along with comprehensive study results presented in the supplementary materials of this paper.
Another concern is the degree to which our results can be generalized to and across the query plan representations of other DBMSs. We selected representative DBMSs of various data models: relational, document, graph, and time series.
The last concern is the degree to which the study really assesses the research questions we aim at. We collected the data ourselves, and our analysis of the semantics may be inconsistent with the intentions of the developers that implemented the query plans.
\section{Related Work}

\emph{Improving query optimization.} Query optimization is a long-standing challenge and a core database research topic with a significant body of related work. We discuss several representative works here.  Ioannidis \etal~\cite{ioannidis2003history} proposed to sample data in single columns for estimating the cardinality of a query plan. Ilyas \etal~\cite{ilyas2004cords} proposed to detect correlations across columns, which enables a multi-column sampling~\cite{poosala1997selectivity} for a more accurate estimation of cardinality. Lakshmi \etal~\cite{lakshmi1998selectivity} proposed using machine learning to estimate cardinality. Estimated cardinality, as well as other estimated costs, are considered in cost models. Babcock \etal~\cite{babcock2005towards} proposed cost distribution to choose a query plan with the lowest cost in a robust way. Apache published a unified framework Calcite~\cite{begoli2018apache} to replace existing query optimizers. Compared to them, we studied various query plan representations, rather than improving query optimization.

\emph{Benchmark and empirical study of query optimization.} Various papers studied query optimizations to understand their performance through empirical studies. Leis \etal~\cite{leis2015good} experimentally studied the contributions of query optimization components to overall performance. Ortiz \etal~\cite{jennifer2019empirical} provided an empirical analysis on the accuracy, space, and time trade-off across several machine learning algorithms for cardinality estimation specifically. Harmouch \etal~\cite{harmouch2017cardinality} conducted an experimental survey on cardinality
estimation, focusing on estimating the number of distinct values. While we also present an empirical study in the context of DBMSs, we focus on query plan representations, rather than the performance, algorithms, and internal components of query optimization.

\emph{Applications based on serialized query plans.} Several applications based on serialized query plans exist. QE3D~\cite{scheibli2015qe3d} visualizes distributed serialized query plans for an intuitive understanding and analysis. Machine learning algorithms have utilized serialized query plans for query optimization. Yuan \etal~\cite{yuan2020automatic} used machine learning to select optimal views. Yu \etal~\cite{yu2020reinforcement} used reinforcement learning to determine the join order. Marcus \etal~\cite{marcus2022bao} and Ryan \etal~\cite{marcus2019neo} applied machine learning algorithms to generate query plans. Zhao \etal~\cite{zhao2022queryformer} used machine learning to convert serialized query plan representations to vector representations to facilitate other machine learning algorithms. In this work, we propose a unified query plan representation, which reduces the effort to build the applications based on serialized query plans.

\emph{Standardization.} Multiple works were proposed to standardize DBMSs. The most related work is Substrait~\cite{substrait}, which proposes a unified representation for DBMS-agnostic relational algebra, while we studied DBMS-specific query plans. Feng \etal~\cite{feng2012towards} proposed a unified architecture to reduce the implementation effort for in-RDBMS analytics. 
Mitschang~\cite{mitschang1988towards} proposed a unified view of design data and knowledge representation when supporting database systems to non-standard applications, such as Computer-Aided Design (CAD). Ginsburg \etal~\cite{ginsburg1992pattern} proposed a unified approach to query sequenced data. 
Gueidi \etal~\cite{gueidi2021towards} proposed a unified modeling method for Non-relational DBMSs (NoSQL) to facilitate the applications on NoSQL. Compared to these works, we propose a unified framework for the query plan representations in DBMSs.

\section{Conclusion}
We have presented an exploratory case study to investigate how query plan representations are in nine widely-used DBMSs. Our study has shown that query plan representations share conceptual components among different DBMSs: operations, properties, and formats. Based on the study, we designed the unified query plan representation to reduce the effort to build applications based on query plans. We implemented a reusable library \method to maintain the unified representation, and evaluated it on five DBMSs. The results show that existing testing methods can be efficiently adopted, finding \numbugs previously unknown and unique bugs. Additionally, existing DBMS-specific visualization tools could support at least five DBMSs by using \method with only moderate implementation effort, and \method enables comparing query plans in different DBMSs, which provides actionable insights.
This paper provides a comprehensive study of query plan representations, and can be used as a reference for other research on query plans. We believe the unified query plan representation provides more opportunities to research serialized query plans in the future.

\section*{Acknowledgments}
This research is supported by the National Research Foundation, Singapore, and Cyber Security Agency of Singapore under its National Cybersecurity R\&D Programme (Fuzz Testing). Any opinions, findings and conclusions, or recommendations expressed in this material are those of the author(s) and do not reflect the views of National Research Foundation, Singapore, and Cyber Security Agency of Singapore.

\bibliographystyle{IEEEtran}
\bibliography{references}

\end{document}